\documentclass[fleqn,twoside]{article}
\usepackage{espcrc2}
\usepackage{graphicx}
\usepackage[figuresright]{rotating}

\newcommand{\AmS}{{\protect\the\textfont2
  A\kern-.1667em\lower.5ex\hbox{M}\kern-.125emS}}

\title{The Thermal Width of the Glueball at Non-Zero Temperature}

\author{Noriyoshi Ishii\address{Radiation Laboratory,
	The Institute of Physical and Chemical Research (RIKEN),\\
	2-1 Hirosawa, Wako, Saitama 351-0198, JAPAN}%
	\thanks{The  lattice  calculations  have  been  performed
	on NEC-SX5  at  Osaka University.},
	Hideo Suganuma\address{Faculty of Science,
	Tokyo Institute of Technology,\\
	2-12-1 Ohkayama, Meguro, Tokyo 152-8552, JAPAN},
	Hideo Matsufuru\address{Yukawa Institute for Theoretical Physics,
	Kyoto University,\\
	Kitashirakawa-Oiwake, Sakyo, Kyoto 606-8502, JAPAN}
}

\newcommand{\Eq}[1]{Eq.~({\ref{#1}})}
\newcommand{\Fig}[1]{Fig.\protect\ref{#1}}

\newcommand{\Tate}{\rule{0cm}{1.1em}}
\newlength{\Tatescale} 

\newcommand{\Bs}{\hspace*{-0.5em}}

\newlength{\figwidth} 
\begin{document}
\begin{abstract}
We use  SU(3) anisotropic lattice QCD  at quenched level  to study the
$0^{++}$  glueball  correlator  at  various  temperature  taking  into
account  the   possible  existence  of   the  thermal  width   in  the
ground-state peak.  For this purpose, we adopt the Breit-Wigner ansatz
for the  appropriate fit-function for  the lattice data  obtained with
5,500--9,900 gauge  configurations at each  $T$. The results  show the
significant  thermal width  broadening as  $\Gamma(T_c) \sim  300$ MeV
with a reduction in the  peak center as $\Delta\omega_0(T_c) \sim 100$
MeV near the critical temperature $T_c$.
\end{abstract}
\maketitle
\section{INTRODUCTION}
At  finite  temperature/density,  the  vacuum properties  of  QCD  are
expected to  change such as the  reduction of the  string tension, the
partial chiral restoration, etc.
Such  changes of  the  vacuum  properties should  be  followed by  the
changes  of  the  hadron   properties,  since  hadrons  are  composite
particles consisting of quarks and gluons.
The  hadronic pole-mass  shifts are  thus considered  to serve  as the
important precritical  phenomena of the QCD phase  transition near the
critical  temperature $T_c$,  and  were extensively  studied by  using
various      QCD-motivated      low-energy     effective      theories
\cite{hatsuda,miyamura,hatsuda2,ichie}.  These  studies suggested  the
pole-mass reductions  of charmoniums, light $q\bar{q}$  mesons and the
glueball near the critical temperature.
Recently, the  accurate pole-masses measurement  at finite temperature
with  the lattice  QCD became  possible  by means  of the  anisotropic
lattice \cite{klassen,taro}.   Quenched-level Monte Carlo calculations
showed  that  the pole-masses  of  the  $q\bar{q}$  mesons are  almost
unchanged from their zero-temperature  values in the confinement phase
\cite{taro,umeda}, while the pole-mass  of the $0^{++}$ glueball shows
the $300$ MeV reduction near the critical temperature \cite{ishii}.
These tendencies are consistent with the recent lattice studies on the
screening mass at finte temperature \cite{laermann,gupta}.
In  these   analysis,  the  bound-state   peaks  are  assumed   to  be
sufficiently narrow.   However, each bound-state  acquires the thermal
width  through the  interaction with  the heat  bath, and  the thermal
width is  expected to grow up  with temperature, which may  leads to a
possible  collapse of the  narrow-peak assumption  in some  cases.  In
this paper, we  first discuss what is the  expected consequence of the
thermal width  broadening.  Then,  we propose the  Breit-Wigner ansatz
for   the  fit-function   for  the   temporal  correlator   at  finite
temperature. We finally show  the results of the Breit-Wigner analysis
of   the   $0^{++}$   glueball   correlator  at   finite   temperature
\cite{ishii2}.
\section{THE BREIT-WIGNER ANSATZ}
We    consider    the     temporal    correlator    $G(\tau)    \equiv
Z(\beta)^{-1}\mbox{Tr}\left(                e^{-\beta               H}
\phi(\tau)\phi(0)\right)$. Its spectral representation is given as
\begin{eqnarray}
	G(\tau)
=
	\int_{-\infty}^{\infty}
	{d\omega \over 2\pi}
	{
		\cosh\left(\omega(\beta/2 - \tau)\right)
	\over
		2\sinh(\beta\omega/2)
	}
	\rho(\omega),
\label{spectral.representation}
\end{eqnarray}
where    $H$   denotes    the    QCD   Hamiltonian,    $Z(\beta)\equiv
\mbox{Tr}(e^{-\beta H})$  the partition function,  $\phi(\tau)$ is the
zero-momentum projected glueball  operator \cite{ishii} represented in
the  imaginary-time Heisenberg picture  as $\phi(\tau)  \equiv e^{\tau
H}\phi(0)e^{-\tau  H}$.   Here,  $\rho(\omega)$ denotes  the  spectral
function
\begin{eqnarray}
	\rho(\omega)
&\equiv&
	\sum_{m,n}
	{ |\langle n|\phi|m \rangle |^2 \over Z(\beta) }
	e^{-\beta E_m}
\\[-0.5em]\nonumber
&&	\times 2\pi
	\left(\Tate
		\delta(\omega - \Delta E_{nm})
	-	\delta(\omega - \Delta E_{mn})
	\right),
\label{spectral.function}
\end{eqnarray}
where $E_n$  denotes the energy  of $n$th excited states,  and $\Delta
E_{mn} \equiv E_m - E_m$.  Note that $\rho(\omega)$ is odd in $\omega$
reflecting  the  bosonic nature  of  the  glueball.   By adopting  the
appropriate ansatz for $\rho(\omega)$, we can extract various physical
quantities such  as the pole-mass  and the width through  the spectral
representation \Eq{spectral.representation}.

We  first consider  the  case  where the  bound-state  peak is  narrow
\cite{taro,umeda,ishii}.    In   this   case,   by   introducing   the
temperature-dependent   pole-mass   $m(T)$,   $\rho(\omega)$  can   be
parameterized as
\begin{equation}
	\rho(\omega)
\simeq
	2\pi A
	\left(\Tate
		\delta(\omega - m(T))
	-	\delta(\omega + m(T))
	\right),
\label{single.pole}
\end{equation}
where  $A$  represents the  strength.   The  second delta-function  is
introduced  to  respect  the  odd-function nature  of  $\rho(\omega)$.
Since  the  corresponding $G(\tau)$  reduces  to  a single  hyperbolic
cosine,  the  pole-mass  measurement  at  finite  temperature  can  be
performed in  the same  way as the  standard mass measurement  at zero
temperature \cite{morningstar}.

We next  consider the case where  the thermal width is  wide.  In this
case,  the peak  center  $\omega_0$ of  $\rho(\omega)$ represents  the
observed ``mass'' of the thermal hadron.  What follows the narrow-peak
assumption now  ?  To consider this,  we notice that  $G(\tau)$ can be
thought of as a weighted average of hyperbolic cosines with the weight
as
\begin{equation}
	W(\omega)
\equiv
	\frac{
		\rho(\omega)
	}{
		2\sinh(\beta\omega/2)
	}.
\end{equation}
Here, $2\sinh(\beta\omega/2)$  in the denominator works  as the biased
factor, which  enhances the smaller $\omega$  region while suppressing
the  larger  $\omega$  region.
Consequently,  the pole-mass  $m(T)$, which  is approximated  with the
peak  position  of  $W(\omega)$,  is  smaller  than  the  peak  center
$\omega_0$ of the spectral function $\rho(\omega)$, i.e., the observed
hadron ``mass'' \cite{ishii2}.

What is the appropriate functional form of the fit-function ?  To find
this,  we consider the  retarded Green  function $G_{\rm{R}}(\omega)$.
At $T=0$, bound-state poles of $G_{\rm{R}}(\omega)$ are located on the
real $\omega$-axis.  At $T > 0$, bound-state poles are moving into the
complex  $\omega$-plane with increasing  temperature.  Suppose  that a
bound-state pole is located at $\omega = \omega_0 - i\Gamma$ as
\begin{equation}
	G_{\rm{R}}(\omega)
=
	{A \over \omega - \omega_0 + i\Gamma} + \cdots,
\end{equation}
where $A$  represents the  residue at the  pole, and  ``$\cdots$'' the
non-singular terms  around the pole.   Since the spectral  function is
the imaginary part of the retarded Green function, the contribution of
this complex pole is expressed in the form of Lorentzian as
\begin{eqnarray}
	\rho(\omega)
&=&
	- 2 \mbox{Im}\left(\Tate G_{\rm{R}}(\omega)\right).
\label{single.peak}
\\\nonumber
&\simeq&	
	2\pi A
	\left(\Tate
		\delta_{\Gamma}(\omega - \omega_0)
	-	\delta_{\Gamma}(\omega + \omega_0)
	\right),
\end{eqnarray}
where     $\displaystyle    \delta_{\epsilon}(x)     \equiv    {1\over
\pi}\mbox{Im}\left(  {1\over x -  i\epsilon} \right)  = {1  \over \pi}
{\epsilon \over  x^2 + \epsilon^2}$  is a smeared  delta-function with
the  width $\epsilon  > 0$.  The  second term  in \Eq{single.peak}  is
introduced to respect the odd-function nature of $\rho(\omega)$.
In   the   limit   $\Gamma\to   +0$,   \Eq{single.peak}   reduces   to
\Eq{single.pole}.
We  thus  see that  the  appropriate  fit-function,  which takes  into
account the  effect of  the non-zero thermal  width, is  the following
Breit-Wigner type as
\begin{equation}
	g(\tau)
\equiv
	\Bs
	\begin{array}[t]{l} \displaystyle
	\int_{-\infty}^{\infty}
		{d\omega \over 2\pi}
		{
			\cosh\left(\omega(\beta/2 - \tau)\right)
		\over
			2\sinh(\beta\omega/2)
		}
	\\\displaystyle
	\times
		2\pi \widetilde A
		\left( \Tate
			\delta_{\Gamma}(\omega - \omega_0)
		-	\delta_{\Gamma}(\omega + \omega_0)
		\right),	
	\end{array}
\label{breit-wigner}
\end{equation}
where  $\widetilde  A$,  $\Gamma$  and $\omega_0$  are  understood  as
fit-parameters, corresponding  to the  residue, the thermal  width and
the peak center, respectively.
Note  that  $g(\tau)$  is  a  generalization of  the  ordinary  single
hyperbolic cosine fit-function.
In order to use Eqs.(\ref{single.pole}) and (\ref{single.peak}), it is
essential  to suppress  the  higher spectral  contributions.  This  is
usually achieved by appropriate choices  of the fit-ranges and also by
improving  the  glueball operator,  for  instance,  with the  smearing
method \cite{ishii2}.
\section{NUMERICAL RESULT}
We   use   the  SU(3)   anisotropic   lattice   plaquette  action   as
$\displaystyle
	S_{\rm{G}}
=
	{\beta_{\rm{lat}}\over N_c}{1\over\gamma_{\rm{G}}}
	\sum_{s,i<j\le   3}
	\mbox{Re}\mbox{Tr}( 1  - P_{ij}(s) )
+
	{\beta_{\rm{lat}}\over    N_c}\gamma_{\rm{G}}
	\sum_{i,j\le  3}
	\mbox{Re}\mbox{Tr}( 1 -  P_{i4}(s) )
$,  where  $P_{\mu\nu}(s)  \in  {\rm{SU(3)}}$  denotes  the  plaquette
operator in the $\mu$-$\nu$-plane.  The lattice parameter and the bare
anisotropic parameter are  fixed as $\beta_{\rm{lat}}\equiv 2N_c/g^2 =
6.25$ and $\gamma_{\rm{G}} = 3.2552$, respectively, so as to reproduce
renormalized anisotropy $\xi  \equiv a_s/a_t = 4$ \cite{klassen,taro}.
These  parameter  set  reproduces   $a_s^{-1}  =  2.341(16)$  GeV  and
$a_t^{-1} =  9.365(66)$ GeV, where  the scale unit is  introduced from
the on-axis data  of the static inter-quark potential  with the string
tension  $\sqrt{\sigma}  =   440$  MeV.   Numerical  calculations  are
performed on  the lattice of the  size $20^3 \times  N_t$ with various
$N_t$.  The  critical temperature $T_c$  on this lattice  is estimated
from the  behavior of the Polyakov-loop susceptibility  as $T_c \simeq
280$ MeV. The pseudo-heat-bath algorithm  is adopted for the update of
the gauge configurations.  In order to construct the temporal glueball
correlators,  we  use  5,500   to  9,900  gauge  configurations.   The
statistical data are  divided into bins of the size  100 to reduce the
possible  auto-correlations   near  the  critical   temperature.   The
smearing  method is  used to  obtain the  improved  glueball operator,
which  is   determined  by  examining  its  behavior   at  the  lowest
temperature, i.e., $T=130$ MeV.

\begin{figure}
\includegraphics[width=\figwidth]{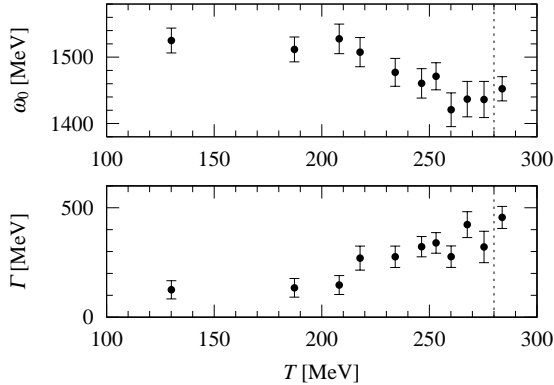}
\vspace{-1cm}
\caption{The center  $\omega_0$ and the thermal width  $\Gamma$ of the
lowest  $0^{++}$ glueball  peak plotted  against temperature  $T$. The
vertical  dotted lines  indicate the  critical  temperature $T_c\simeq
280$ MeV.}
\label{figure}
\end{figure}
The peak center and the thermal width of thermal glueball are obtained
by the  best-fit analysis of the suitably  smeared glueball correlator
$G(\tau)/G(0)$  with \Eq{breit-wigner}  at  various temperatures.   In
\Fig{figure},  the  peak  center  $\omega_0$  and  the  thermal  width
$\Gamma$  are plotted against  temperature.  While  narrow-peak ansatz
leads  to   the  pole-mass  reduction   of  300  MeV  near   $T_c$  in
Ref.\cite{ishii},   the  Breit-Wigner   analysis  indicates   a  small
reduction  in  the peak  center  as  $\Delta  \omega_0(T_c) \sim  100$
MeV.  Instead, we observe  a significant  thermal width  broadening as
$\Gamma(T_c) \sim 300$ MeV.
\section{SUMMARY}
We have  studied the temporal  $0^{++}$ glueball correlator  at finite
temperature using SU(3) anisotropic lattice QCD at quenched level with
5,500 to 9,900 gauge configurations at each temperature.
We have proposed the Breit-Wigner  ansatz for the fit-function to take
into  account the  effect  of  the non-zero  thermal  width at  finite
temperature.
We  have applied the  Breit-Wigner analysis  to the  temporal glueball
correlator at finite temperature, and have observed a slight reduction
of  the peak center  as $\Delta  \omega_0(T_c) \sim  100$ MeV  and the
significant broadening of the  thermal width as $\Gamma(T_c) \sim 300$
MeV in the vicinity of the critical temperature $T_c$.

\end{document}